\documentclass[preprint,12pt]{elsarticle}

\usepackage{amssymb}

\journal{Comptes Rendus Physique}

\begin{document}

\begin{frontmatter}



\title{A review of Monte Carlo simulations for the Bose-Hubbard model with diagonal disorder}


\author{Lode Pollet}

\address{Department of Physics, Arnold Sommerfeld Center for Theoretical Physics and Center for NanoScience, University of Munich, Theresienstrasse 37, 80333 Munich, Germany}

\begin{abstract}
We review the physics of the Bose-Hubbard model with disorder in the chemical potential focusing on recently published analytical arguments in combination with quantum Monte Carlo simulations. Apart from the superfluid and Mott insulator phases that can occur in this system without disorder, disorder allows for an additional phase, called the Bose glass phase. The topology of the phase diagram is subject to strong theorems proving that the Bose Glass phase must intervene between the superfluid and the Mott insulator and implying a Griffiths transition between the Mott insulator and the Bose glass. The full phase diagrams in 3d and 2d are discussed, and we zoom in on the insensitivity of the transition line between the superfluid and the Bose glass in the close vicinity of the tip of the Mott insulator lobe. We briefly comment on the established and remaining questions in the 1d case, and give a short overview of numerical work on related models. 
\end{abstract}


\end{frontmatter}


\section{Introduction}
\label{sec:introduction}

The interplay between disorder and interactions is a subtle and long-standing problem in condensed matter physics, especially for bosons on a lattice. In the absence of interactions, the smallest amount of disorder will lead to Anderson localization of the atoms when all bosons occupy the same lowest-energy localized state. This limit is clearly pathological. In the absence of disorder, a sufficiently strong interaction will drive the system towards a gaped and incompressible Mott insulator provided the density is commensurate. Although both disorder and interaction tend to localize the bosons, the superfluid phase is quite robust against disorder. This compressible and gapless superfluid is found over a large part in the phase diagram: it exists for weak interactions, but also  for interactions where the pure system is an insulator so-called disorder induced superfluidity is possible. For even stronger disorder and/or interactions it will ultimately go over in a new phase, the Bose glass phase~\cite{Giamarchi1987, Giamarchi1988}. This is a gapless and compressible but insulating phase. Its properties thus defy intuitive notions about conductivity based on Fermi liquid theory. Fisher {\it et al.}, building on the one-dimensional work by Giamarchi and Schulz~\cite{Giamarchi1987, Giamarchi1988}, argued the existence of the Bose glass phase in any dimension and gave an extensive description of its properties~\cite{Fisher1989}.

For a long time, there was controversy whether a direct transition between a superfluid (SF) and a Mott insulator (MI) in the presence of disorder was possible~\cite{Freericks1996,Scalettar1991,Krauth1991,Zhang1992,Singh1992,Makivic1993, Wallin1994,Pazmandi1995,Pai1996,Svistunov1996,Pazmandi1998,Kisker1997, Herbut1997,Sen2001, Prokofev2004, Wu_Phillips2008,Bissbort2009,Weichman2008_1,Weichman2008_2}. Fisher {\it et al.} argued that this was unlikely, though not fundamentally impossible~\cite{Fisher1989}. Curiously, a large number of direct numerical simulations and some approximate approaches observed the unlikely scenario. 
In the past 5 years, a number of theorems have been published that allow to answer these issues and prove the conjectures by Fisher {\it et al.}~\cite{Fisher1989}. Here, we review these theorems, and discuss their influence on the numerically obtained phase diagrams of the disordered Bose-Hubbard model with disorder in the chemical potential by large scale path integral Monte Carlo simulations with worm-type updates.

The main results and the contents of the paper are as follows. To set the ideas, we specify our Hamiltonian and the disorder model in Sec.~\ref{sec:model} with the definition of the phases of interest (superfluid (SF), Mott insulator (MI) and Bose glass (BG)). We proceed in Sec.~\ref{sec:theory} with a brief overview of what is known analytically about such systems. We will introduce Lifshitz regions, analyze the regime of weak interactions, review a sufficient condition by Fisher {\it et al.} for the presence of a BG phase in the vicinity of the MI - BG transition, review the theorem of inclusions stating that a glassy phase must intervene between a gaped and gapless phase in the presence of disorder, which in turn shows that the sufficient condition by Fisher is also necessary and leads to a Griffiths type transition between the BG and the MI. We then move on to the numerical section (see Sec.~\ref{sec:QMC}) containing a brief discussion of path integral Monte Carlo with worm-type updates, and a presentation of the phase diagrams of the disordered Bose-Hubbard Hamiltonian in $d=3,2$ and 1 dimensions. We briefly mention a number of experimental (see Sec.~\ref{sec:experiment}) and other recent numerical studies in Sec.~\ref{sec:other}, before concluding in Sec.~\ref{sec:conclusion}.

\section{Bose-Hubbard model}
\label{sec:model}

\begin{figure}
\includegraphics[angle=-90,width=0.8\columnwidth]{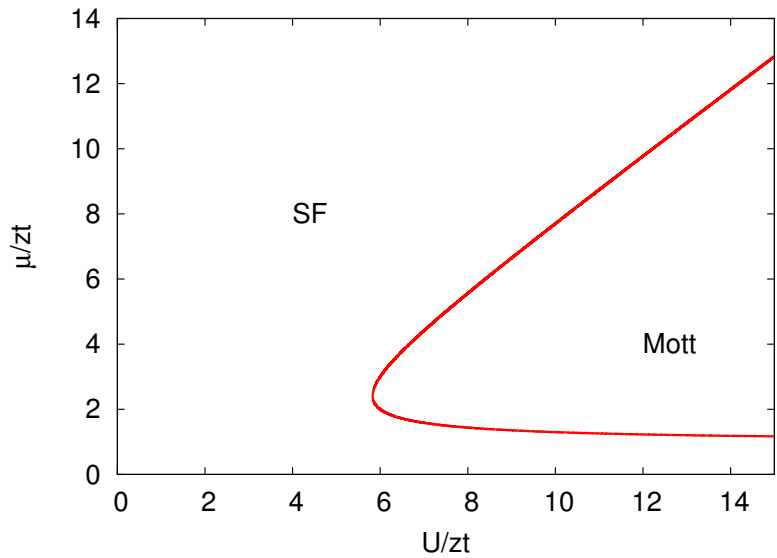}
\caption{ (Color online). Mean-field zero temperature phase diagram  in the $(U, \mu)$ plane in units of $zt$ (where $z$ is the coordination number) of the Bose-Hubbard model without disorder. At commensurate densities, Mott insulators (MI) are formed for strong enough interactions, while a superfluid (SF) is found elsewhere. In this approximation, the tip of the Mott lobe is found for $U/zt \approx 5.83$ where the Mott boundary has a vertical slope~\cite{Fisher1989, Sachdev99}.
}
\label{fig:Mott_MF}
\end{figure}

We consider the Bose-Hubbard model~\cite{Fisher1989} describing scalar bosons on a lattice,
\begin{equation}
H =  -t \sum_{\langle i,j \rangle} b_i^{\dagger}b_j + \frac{U}{2} \sum_i n_i(n_i-1) - \mu \sum_i n_i.
\label{eq:BoseHubbard}
\end{equation}
The operators $b_j$ annihilate a scalar boson on site $j$ while $b_i^{\dagger}$ creates such a boson on site $i$ and is the Hermitean conjugate operator of $b_i$.  These operators satisfy the bosonic commutation relations, $[ b_i, b_j^{\dagger}] = \delta_{i,j}$ and this yields zero for commutators with two creation or two annihilation operators. The
operator $n_i = b_i^{\dagger}b_i$ counts the number of bosons on site $i$.
The first term describes the hopping of bosons between neighboring sites with tunneling amplitude $t$, which we set as our unit $t=1$.
The second term describes the on-site repulsion with strength $U$, while the last term is proportional to the chemical potential $\mu$.
The lattice spacing is set to unity. Unless otherwise specified we have a lattice in mind of linear size $L$ and volume $L^d$ where $d$ is the dimension and we use periodic boundary conditions.
The Bose-Hubbard model is the simplest model that describes a conductor-insulator transition for bosons. It  has three phases, which can easily be identified in the limiting cases~\cite{Fisher1989, Sachdev99}:
First, in the limit of high temperature, the system is a normal liquid.  At zero temperature and $t=0$, the system is a Mott insulator with fixed, integer density, a gap, and zero compressibility. For finite hopping, stable Mott lobes around the $t=0$ insulators are found, which are surrounded by a gapless, compressible superfluid. The superfluid phase also exists at finite temperature. The zero-temperature phase diagram in the mean-field approximation (also known as the decoupling approximation) is shown in Fig.~\ref{fig:Mott_MF}. The true phase diagrams in 3d and 2d look very similar with only quantitative changes especially in the vicinity of the tip of the Mott lobe, which are shifted to lower interaction strengths by about $20\%$ and $30\%$, respectively. In 1d, the tip of the lobe has a cusp reflecting the Kosterlitz-Thouless transition where the shape of the Mott insulator also bends down. There is thus reentrant behavior in 1d when following a line of constant $\mu$ close to the tip of the lobe.

In this review  disorder will be introduced at the level of the chemical potential by $\mu \to \mu_i = \mu - \epsilon_i$, where $\epsilon_i$ is the disordered on-site potential. This is known as diagonal disorder, unlike disorder introduced in the hopping which is known as off-diagonal disorder. We can take $\epsilon_j$ as independent random variables distributed according to the probability density $p(\epsilon/\Delta)$, which satisfies the normalization condition $\int_{-1}^{1} du \, p(u)=1$. The first moment can always be taken to be zero, $\int_{-1}^{1} du \, u p(u)=0$, by absorbing it in the definition of $\mu$. The disorder distribution is taken to be  bounded (otherwise the Mott insulator does not exist), that is $p(u)=0$ if $\vert u \vert > 1$. Formally, the disorder bound $\Delta$ and the disorder distribution dispersion $\delta = \sqrt{ \langle \epsilon^2 \rangle - \langle \epsilon \rangle^2}$ are independent parameters. For the most common
choice of the uniform distribution $p(u)={\rm const}$ (also used in our numerical simulations),
we have $\delta^2=\Delta^2/3$. For the theoretical discussion, however, we allow more general distributions. The disorder can completely be characterized by specifying all its moments. One may also add parameters which control correlations between the potentials on different lattice sites. All these parameters are denoted by $\xi$, and we identify them with the definition of a particular model of disorder. 

With this kind of disorder, a new phase is possible: the Bose glass phase~\cite{Giamarchi1987, Giamarchi1988}, which is an insulator, though gapless and compressible.
While the superfluid is self-averaging, the Bose glass phase is not; that is, most realizations of the superfluid have the same superfluid stiffness in the thermodynamic limit, while for a large though finite system size in the Bose glass phase this property will be absent.
With other types of disorder, other phases such as a Mott glass~\cite{Orignac2001, Prokofev2004} may exist (the Mott glass is incompressible, and can occur for instance for hard-core bosons at half filling with disorder in the hopping, featuring particle-hole symmetry), but we will not consider such cases in this paper.

\section{Analytical aspects}
\label{sec:theory}

In this section we review some old and some recent theoretical arguments that put very strong contraints on the phase boundaries of the disordered Bose-Hubbard model.
We start with a standard textbook discussion of Lifshitz tails.

\subsection{Lifshitz tails}\label{sec:Lifshitz}

In disordered systems, localized states with an energy close to the band edge $E_{\rm min} = -\Delta - 2dt$ play a crucial role. The probability $P$ of finding a state with energy $\delta E = E - E_{\rm min}$ close to $E_{\rm min}$ extended over $\ell$ states is proportional to $(1/\Delta)^{\ell}$. The linear size of such a state is $R \sim \ell^{1/d} r_0$ with $r_0$ a typical length scale such as the lattice constant. Its energy is then $\delta E = t \ell^{-2/d} = t (R / r_0)^2$. From this it follows that the probability $P$ is given by
\begin{equation}
P \sim e^{ - (t / \delta E)^{d/2} c},
\end{equation}
with $c$ a constant. Such Lifshitz regions are thus exponentially rare but can be arbitrarily large (such that the quantization energy can be very low). They pose a challenge for theoretical treatments of disordered systems and will show up several times in the discussions below.

\subsection{Weak interactions}\label{sec:weak_interaction}

This regime was analyzed in detail by Falco {\it et al.}~\cite{Falco2009} and Lugan {\it et al}~\cite{Lugan2007}, and we reproduce here the dependence of the critical line on $U/t$ and $\Delta/t$ by a derivation based on the central limit theorem.
When $U \ll 1$ and $\Delta \ll 1$, the transition between the BG and the SF happens through percolation between localized states (with large localization length) at high energy $E$.  The precise form of the disorder distribution is then irrelevant, and we can use a delta-correlated, Gaussian distribution $V$ with mean 0 and variance $\Delta^2$ satisfying $\langle V(x) V(y) \rangle = \Delta^2 a^d \delta(x-y)$ with $a$ the lattice constant. The requirement of commensurability of the density $n$ is likewise unimportant since Mott physics does not enter into the picture.
The density of (localized) states with a high energy $E$ is the probability that the average potential energy over a large region of linear size $R$ is $E$. This leads to the following estimate for the density of states, using the central limit theorem,
\begin{eqnarray}
\rho(E) & = & \langle \delta \left( \int \frac{ d^dx}{R^d} V(x) - E \right) \rangle \nonumber \\
{} & = & \int d\phi e^{-i \phi E} \langle e^{i \phi \int \frac{ d^dx}{R^d} V(x) } \rangle \nonumber \\
{} & = & \int d \phi e^{-i \phi E} e^{- \frac{\phi^2}{2} \int \int  \frac{ d^dx}{R^d}  \frac{ d^dy}{R^d} \langle V(x) V(y) \rangle} \nonumber \\
{} & \propto & e^{ - \frac{E^2 R^d}{2 \Delta^2 r_0^d}}. 
\end{eqnarray}
Since $E \propto t \frac{r_0^2}{R^2}$ it follows that $\rho(E) \propto e^{-\sqrt{\frac{E^{4-d}t^d}{2 \Delta^4}  }}$. The chemical potential parameter can be estimated as $nU$ in this regime, setting the scale for the critical $E$ when the localized states start overlapping,
\begin{equation}
\Delta_c \propto U^{1 - d/4}.
\label{eq:lowU}
\end{equation}

\subsection{A necessary condition for a gaped system}
\label{sec:theorem_Fisher}

Fisher {\it et al.} proved rigorously that if the bound $\Delta$ on the
disorder strength is larger than the half-width of the energy gap $E_{\rm g/2}$ in the
Mott insulator of the disorder free system, then the system is compressible.
This implies that the transition, characterized by a vanishing superfluid density, goes from a SF to a BG and not to a MI whenever
\begin{equation}
\Delta_c > E_{\rm g/2} .
\label{condition}
\end{equation}
The proof of this theorem is a reasoning typical for disordered systems and based on Lifshitz regions. Consider an arbitrarily large region (however rare) where the chemical potential is roughly constant and lower than $-E_{g/2}$. Locally, this system looks like a superfluid. Provided the finite size quantization of the energy due to the large volume is small enough, such regions can be doped (take for instance 2 such regions, then particles can be transfered from the one to the other at no cost). So the whole system can have no gap in the spectrum and is compressible.

Although this proves that the system is certainly not an insulator whenever $\Delta > E_{g/2}$, it is not a necessary and sufficient condition fixing the location of the MI boundary since the argument does not prove that systems with $\Delta < E_{g/2}$ (and $U > U_c(\Delta=0)$) are gaped. It also does not rule out a direct transition between a SF and the MI. Although the latter was conjectured to be very unlikely by Fisher {\it et al.}, it was only rigorously ruled out in 1d by Svistunov~\cite{Svistunov1996} before the theorem of inclusions (valid in any dimension) was proven, which is the topic of the next section.


\subsection{Theorem of Inclusions}
\label{sec:inclusions}

\begin{figure}
\includegraphics[width=0.8\columnwidth]{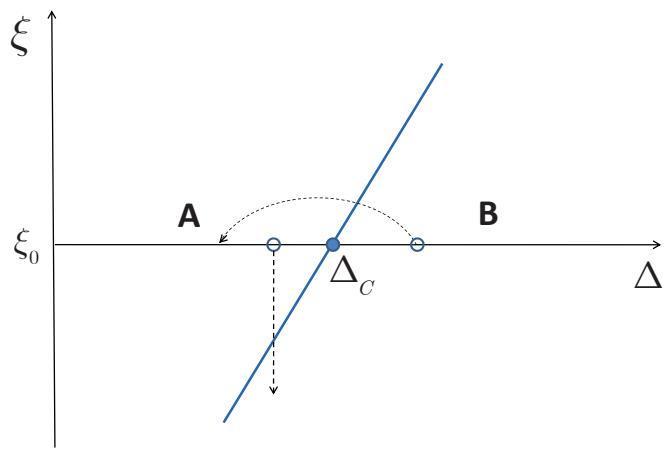}
\caption{ (Color online). A sketch of the generic phase transition line between some phases A and
B in the plane of disorder distribution parameters $\Delta$ and $\xi$, where $\Delta$ is the
bound and $\xi$ is one of the infinite number of parameters characterizing the
disorder distribution function, {\it e.g.}, the variance and its spatial correlations.
Dashed lines with arrows originate from points which determine the disorder properties in the
macroscopic (thermodynamic limit) system, and end at points which characterize disorder
parameters in an arbitrarily large, but finite, domain as a result of rare statistical fluctuations in the same system.
Reprinted figure with permission from Ref.~\cite{Gurarie2009}. Copyright (2009) by the American Physical Society.
}
\label{fig:inclusions}
\end{figure}

This and the next section follow the arguments presented in Refs.~\cite{Pollet2009_disorder, Gurarie2009}.
The theorem of inclusions is explained schematically in Fig.~\ref{fig:inclusions}. Consider a phase transition driven by disorder between two arbitrary phases A and B. 
The curved dashed line between A and B means that inside phase B there exist arbitrarily large lakes that locally look like phase A, since B occurs for stronger disorder than A. This amounts to the same argument as in the previous section. How can we prove the converse, namely that inside A there are arbitrarily large regions that locally look like B?
For a generic disorder model, the transition between A and B will be a function of the disorder bound $\Delta$ as well as of the properties of the disorder distribution specified by $\xi$ (see Sec.~\ref{sec:model}).
Close to the transition line, there exist then arbitrarily large domains in phase A that look like phase B but for a model with a slightly different disorder distribution (and it is always possible to find such a distribution since there are an infinite number of continuous parameters determining the disorder properties), as is explained in the figure by the vertical dashed line with an arrowhead.  

The consequence of the above argument is that it is not possible for phase A to be gaped if phase B is gapless since the domains of arbitrarily large size that look like phase B inside phase A make phase A gapless. As a consequence, no direct transition between the gaped Mott insulator phase and the gapless superfluid phase is possible, but a gapless Bose glass must always intervene in any dimension. The compressibility of the BG phase follows immediately from the existence of rare but arbitrarily large regions where the energy quantization is low enough such that the chemical potential can be slightly varied away from its equilibrium value (esentially the disorder average over the region) without violating the disorder bound. This fails for a Mott glass however.

\subsection{Griffiths transitions}\label{sec:griffiths}

Another consequence of the theorem of inclusions is that it seems to rule out the transition between the MI and the BG at first sight. 
The paradox is resolved by considering the exception implied by the rule: the argument of the previous section does not work when the critical point $\Delta_c$ cannot be identified with a generic case, {\it i.e.,} that $\Delta_c$ is not dependent on $\xi$. A transition which depends only on the bound $\Delta$ cannot be linked to any local physics because when the variance of the disorder distribution goes to zero, the system becomes indistinguishable from a disorder free system in the thermodynamic limit. The transition mechanism must hence be based on statistical fluctuations which explore the possibility of reaching the disorder bound at all sites on larger and larger scales which mimic a regular pure system in an external field.  For the MI, the external field is the chemical potential which drives the MI to a SF whenever it exceeds the gap for particle or hole creation. In the general case it can be any regular external field whose amplitude scales with $\Delta$. This mechanism is nothing but the conjectured Griffiths type  MI--BG transition when the vanishing of the gap at the critical point
is  due to an infinitesimal concentration of rare regions in which the fluctuation of disorder mimics a {\it homogeneous} chemical potential shift \cite{Fisher1989}.
So, the boundary of the MI is given by $E_{g/2}$, and there is no hope to observe this transition numerically. 

\section{Quantum Monte Carlo simulations}
\label{sec:QMC}

Before presenting the phase diagrams of the disordered Bose-Hubbard model in $d=3,2$ and 1, let us first make a few remarks on how such systems can be studied numerically by path integral Monte Carlo simulations with worm-type updates. We refer to Ref.~\cite{Pollet2012_review} for a recent review of the method with applications for cold gases.

\subsection{The worm algorithm}
\label{sec:worm}

The worm algorithm~\cite{Prokofev1998_worm} is a path integral Monte Carlo method (PIMC) ideally suited for studying systems with superfluidity. Let us first see how the algorithm works in the absence of disorder.
The starting point  is the following decomposition for the partition function,
\begin{equation}
Z = {\rm Tr} e^{-\beta H} = {\rm Tr} {\mathcal{T} }e^{-\beta H_0} \exp \left[ - \int_0^{\beta} d\tau H_1(\tau) \right],
\end{equation}
where $H_1(\tau) = e^{\tau H_0}H_1 e^{-\tau H_0}$.  As basis we use the Fock basis of occupation numbers, $\vert \{ n_i \} \rangle$, in which case the above formula refers to a strong coupling expansion. The potential energy and the chemical potential term are diagonal in this basis, and are combined into $H_0$. The kinetic term ($H_1$), which is a one-body operator, is not diagonal in this basis but will lead to transitions between Fock states with matrix elements $\langle \ldots n_i-1, n_j+1, \ldots \vert -tb_j^{\dagger}b_i \vert \ldots n_i, n_j \ldots \rangle = -t \sqrt{n_i(n_j+1)}$.
The trace is taken over all Fock basis states. The exponential is expanded into a time-ordered product~\cite{Mahan, NegeleOrland, FetterWalecka},
\begin{equation}
Z =  {\rm Tr} {\mathcal{T} }e^{-\beta H_0} \left[ 1   - \int_0^{\beta} d\tau H_1(\tau) +  \int_0^{\beta} d\tau_1 \int_{\tau_1}^{\beta} d\tau_2 H_1(\tau_1) H_1(\tau_2) + \ldots   \right].
\label{eq:BH_expansion}
\end{equation}
The inverse temperature $\beta = 1/T$ is understood as an imaginary time, where the matrix elements of the operators $\exp [-\Delta \tau H_0]$ act as propagators between the different states at $\tau_1, \tau_2, \ldots$, with $0 < \tau_1 < \tau_2 < \ldots < \tau_j < \ldots < \tau_n < \beta$. 
There is a pictorial representation of this expansion formula, in which each $H_1$ term is shown as a kink where the occupation numbers change. The legs of the kinks are connected to other kinks by lines whose properties (can be thickness or multiple lines) represent the occupation number of the propagator connecting the two kinks. At time $\tau = 0$ one can now select a particle and follow its motion in space and imaginary time. One sees that all permutations of the particles can be represented this way since at time $\tau = \beta$ the propagated line either closes on the same particle or on another particle, which is allowed by indistinguishability of the particles. One calls a worldline the trajectory of a particle propagating in space and imaginary time.

The expansion in path integral Monte Carlo  is always understood over a finite volume and finite maximum imaginary time. There is no singularity caused by a phase transition, which can only be studied in a finite size scaling analysis. The expansion is written in terms of an entire function so that there are no non-physical divergencies in the PIMC formulation. 

According to statistical mechanics, the expectation value of an observable $A$ is given by  $\langle A \rangle = \frac{1}{Z} {\rm Tr} A e^{-\beta H}$. In PIMC, a statistical interpretation is given to Eq.~\ref{eq:BH_expansion} by introducing weights $w$ through $Z = {\rm Tr}_{\vert \{n_i \} \rangle} w({\vert \{ n_i \} \rangle})$. We now have to statistically generate all possible configurations according to their weights by generating all possible expansion orders and matrix elements, assign an (unnormalized) weight to each one of them, evaluate the observable $A$ in every configuration, and sum all these contributions. For instance, one can perform local updates by inserting pairs of hopping elements, in which a particle hops from a site to its neighbor, and back at a later time.

At high temperatures or deep in the Mott insulating phase, the kinetic energy is small compared to $U$ and/or $T$, and few perturbation orders are needed in Eq.~\ref{eq:BH_expansion}. Such is not the case when the contributions to the free energy coming from the hopping terms are large: There is no reason to expect that the local updates would be efficient. Even worse, they are not ergodic: The low energy states in a superfluid are given by states with a different winding number. These are paths in which a particle winds around the full linear length of the system before closing on itself again. Configurations with different winding numbers are topologically distinct and cannot be transformed into each other by local updates alone. The winding number $W$ is directly related to the superfluid density through $\rho_s = \frac{\langle W^2 \rangle L^{2-d} } {d \beta}$~\cite{Pollock1987}, with $d$ the dimension of the system. 

The worm algorithm has completely solved those ergodicity problems~\cite{Prokofev1998_worm}. Instead of working with the partition function $Z$ alone one also works in the Green function sector $Z_G$,
\begin{equation}
Z_G = {\rm Tr} \mathcal{T} \{ b_i(\tau_0) b_j^{\dagger}(\tau) e^{-\beta H} \}.
\end{equation}
Pictorially, the operators $b_i(\tau_0)$ $b_j^{\dagger}(\tau)$ are open ends delimiting a segment of a worldline and are called the worm head and the worm tail. The worms can be on any site and any time. A correct algorithm allows for the transition between the Green function sector and the partition function sector (where measurements of observables such as the energy and the superfluid stiffness are done), and allows to move the worms around in configuration space.  Worms also have the ability to insert and remove hopping elements (kinks).
Since the worm operators correspond to open ends on a world line segment they have no problem in exploring configurations with different winding numbers. All updates are local in the Green function sector, which means that all acceptance factors can be made of order unity. In phases with off-diagonal long-range order (either true long-range order  such as seen in a Bose-Einstein condensate or quasi long-range order with correlation functions  decaying as a power-law) the worms will preferentially be far away from each other in configuration space, {\it i.e.}, we are efficient in describing the physics of those phases. In order to distinguish between the superfluid, Bose glass and Mott insulating phases, we need to compute the superfluid stiffness and the compressibility. The latter corresponds to the winding number fluctuations in imaginary time, {\it i.e.} by the fluctuation-dissipation theorem the compressibility is $\kappa = \frac{ \partial n}{\partial \mu} = \beta (\langle N^2 \rangle - \langle N \rangle^2)$. Because the worm algorithm is formulated in the grand-canonical ensemble, the latter quantity is readily accessible in simulations. There exist other algorithms that build on the same physical idea of worms such as the stochastic series expansion algorithm with directed loops~\cite{Sandvik1999, Sandvik2003}, directed loops in path-integral representation~\cite{Pollet2007}, algorithms with multiple worms present (which may lead to slowdowns)~\cite{Rousseau2007, Rousseau2008}, as well as canonical formulations~\cite{Rombouts2006, VanHoucke2006}. Rather than emphasizing minor differences, the important issues are that sectors with different winding numbers must be sampled efficiently (this is the idea behind worms and ensures that autocorrelation times of order unity can be reached everywhere in the phase diagram, at least in theory) as well as an implementation with an efficient data structure (such that the autocorrelation times of order unity can also be reached in practice).

Second, we discuss how the worm algorithm works in the presence of chemical potential disorder. Because such disorder is diagonal in the Fock basis, there are essentially no changes to the above algorithm other than that the chemical potential is now site-dependent. One has to compute thermodynamic observables for a fixed disorder realization, and then average those observables over all possible disorder realizations (by the replica trick). So although we see that the algorithm is essentially unaltered, simulations of disordered systems can still be notoriously difficult and contain a number of pitfalls where the inexperienced (and often also the experienced) practitioner needs to pay attention (most of these considerations not only apply to Monte Carlo simulations, but to any numerical approach):
\begin{enumerate}
\item There is the risk of a lack of ergodicity and inefficient sampling. For the disordered Bose-Hubbard model, the worm algorithm has effectively solved this problem, and this should no longer be a main issue if one checks that the simulations are well converged.
\item One underestimates the number of realizations that are needed. Especially when rare fluctations (Lifshitz regions) have an impact, a too small number of realizations can easily lead to wrong conclusions. Cases are known where $10^5$ realizations are needed in order to have converged answers; otherwise small drifts can be observed. The Bose glass also does not have the property of self-averaging, meaning that the distribution of the stiffness has to become very broad on accessible system sizes, so an accurate description of a broad distribution understandably requires many disorder samples. But also weak and marginal superfluid phases have broad distributions on mesoscopic system sizes. 
\item System sizes may be too small. Griffiths transitions can never be observed on a finite system, and so numerical results should always be checked against the analytical theorems. The compressibility turns out to be a very small number ($10^{-5}$ and smaller was observed numerically) in the vicinity of the Mott lobe, also making a direct observation extremely challenging especially in the vicinity of the tip of the Mott lobe.
\item In cases where the theory is not well established, numerical fluctuations may be hard to control and difficult to interpret. Simulations in one dimension for weak interactions provided a prominent example (see however Sec.~\ref{sec:1d}).
\end{enumerate}

In the next sections we discuss the phase diagrams in 3d, 2d and 1d as found by Monte Carlo simulations.

\subsection{Phase diagram of the disordered Bose-Hubbard model in 3d}
\label{sec:3d}

\begin{figure}
\begin{center}
\includegraphics[width=0.8\columnwidth]{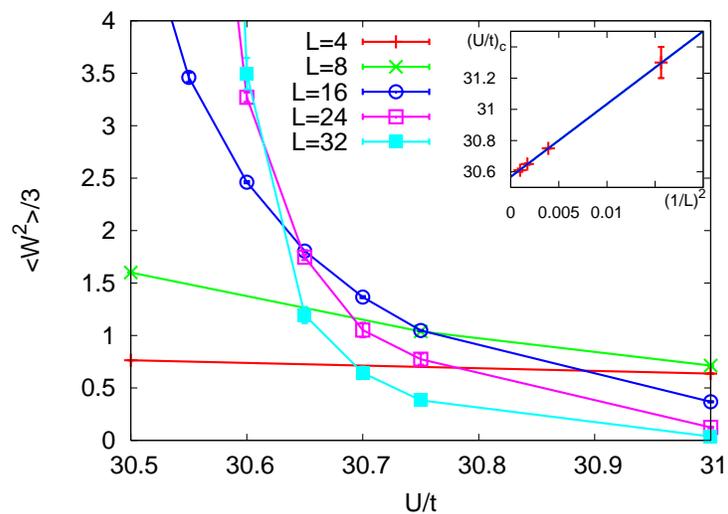}
\caption{ (Color online). Finite size scaling analysis for the winding number fluctuations for $\Delta / t = 5$ using the dynamical exponent $z=2$, chosen for numerical convenience (a calculation with $z=1$ produced the same critical point), and starting value $\beta t = 1$ for $L=4$. The inset shows the extropalation of the intersection points with a $(1/L)^2$ law yielding $(U/t)_c = 30.57(2)$. The figure is a reprint combination of figures from Ref.~\cite{Pollet2009_disorder} with permission. Copyright (2009) by the American Physical Society.
}
\label{fig:FSS}
\end{center}
\end{figure}

\begin{figure}
\begin{center}
\includegraphics[width=0.8\columnwidth]{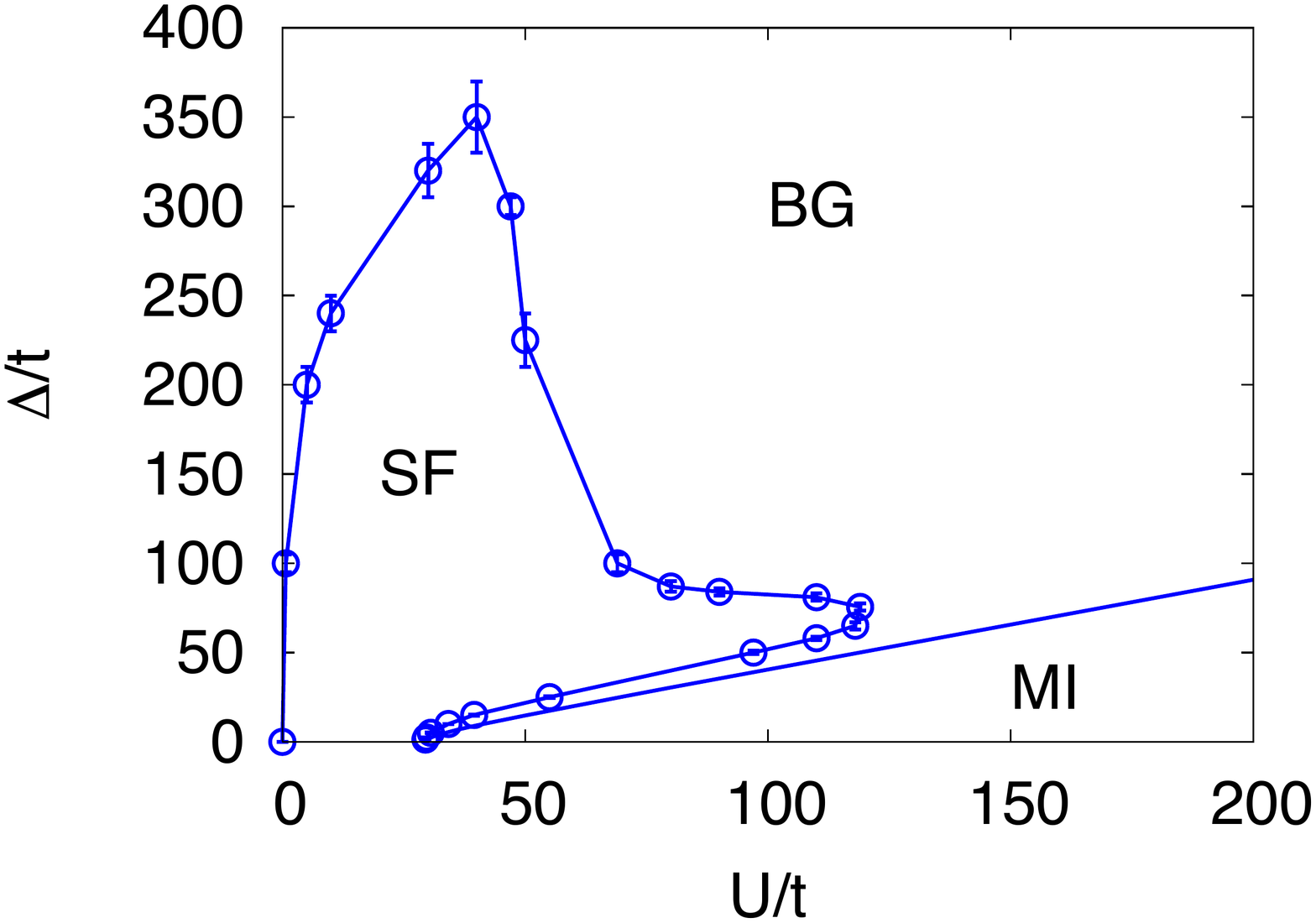}
\caption{ (Color online). Phase diagram of the disordered 3d Bose-Hubbard model at unit filling, obtained by a finite-size analysis of winding numbers. In the absence of disorder, the system undergoes a quantum phase transition between SF and MI phases. The presence of disorder allows for a compressible, insulating BG phase, which always intervenes between the MI and SF phases because of the theorem of inclusions~\cite{Pollet2009_disorder, Gurarie2009}. The transition between MI and BG is of the Griffiths type, as an exception implied by the theorem of inclusions~\cite{Gurarie2009}. At $U/t \to 0$, the SF-BG transition line has an infinite slope~\cite{Falco2009}.
Reprinted figure with permission from Ref.~\cite{Gurarie2009}. Copyright (2009) by the American Physical Society.
}
\label{fig:phasediagram_disorder}
\end{center}
\end{figure}

Critical points on the SF-BG transition line are determined using a finite size scaling analysis of the superfluid density. 
The scaling equation for the superfluid density for a dimensionless detuning $\delta$ reads~\cite{Fisher1989}
\begin{equation}
\rho_s(\delta, L) = \xi^{-(d+z-2)} f_s( \xi / L, \xi^z / \beta).
\end{equation}
On finite system sizes the correlation length $\xi \sim \delta^{-\nu}$ occuring in front of the universal scaling function $f_s$ will be cut off by the system size. For a quantum phase transition, $\beta$ must be scaled with $L^z$ according to the dynamic exponent $z$. Hence, by plotting $\rho_s L^{-(d+z-2)}$ versus detuning for different system sizes $L$ and determining the consecutive crossing points, the critical point can be determined. It will be a single crossing point to leading order by scale invariance if the correct scaling exponents are used.
An equivalent viewpoint is noting that at the critical point the winding number fluctuations, which are integer numbers, must be scale invariant.
 For a disorder free Bose-Hubbard model at the tip of the Mott lobe, $z=1$. In the presence of disorder and when the compressibility $\kappa$ is finite, it was argued that $z=d$ in Ref.~\cite{Fisher1989}. This was however recently questioned in numerical~\cite{Makivic1993, Priyadarshee2006} and analytical ~\cite{Weichman2007} studies.
How the finite size scaling is done in practice, can be seen in Fig.~\ref{fig:FSS}. Because $z=3$ implies a too fast increase of $\beta$ with $L$, we opted for a lower $z=2$ out of convenience. Any $z>0$ is fine to find the critical point. It is clear from Fig.~\ref{fig:FSS} that the critical point on the SF-BG transition line can be determined reliably, which allows an accurate numerical determination of the entire superfluid boundary.

The resulting phase diagram of the 3d Bose-Hubbard model is shown in Fig.~\ref{fig:phasediagram_disorder} for unit density. The BG phase always intervenes between the SF and MI, while the transition from the MI to the BG phase is of the Griffiths type, in agreement with the preceding theoretical discussion. 
The transition line is then determined by measuring the gaps in the disorder-free MI.  At $U/t \to 0$,  the data are consistent with Eq.~\ref{eq:lowU}. Reentrant behavior is also clearly seen in the phase diagram~\cite{Krauth1991, Krauth1991_EPL}, as well as a large region where disorder induced superfluidity is possible ({\it i.e.}, superfluidity is possible with disorder, but does not occur in the disorder free system). The BG phase is connected to the $U=0$ limit described by Anderson localization. The transition temperature inside the superfluid finger is very low, {\it e.g.} $T_c/t = 0.37(5)$  for $\Delta/t = 65$ and $U/t = 60$~\cite{Pollet2009_disorder, Gurarie2009}, and the corresponding condensate fraction at low temperature is equally low.
What is also remarkable are the large scales for the SF-BG transition lines. For intermediate interactions, the superfluid phase extends to $\Delta/t \approx 300$ which at first seems to have little to do with the microscopic parameters of the Hubbard model. It was explained in Ref.~\cite{Gurarie2009} as follows:
Similar as in Sec.~\ref{sec:weak_interaction} we expect that the transition between the SF and the BG is still given by percolation at moderate interaction strength and $\Delta \gg t$, but commensurability plays a role now and the localization length will be of the order of the lattice constant~\cite{Bulka1987}.
From the local Hamiltonian, the density can be computed, and setting it equal to one leads to the condition
\begin{equation}
\mu = -U/2 - \Delta + 2\sqrt{U\Delta}.
\end{equation}
A site will be occupied if its disorder lies within the $(-\Delta, U/2 + \mu) =(-\Delta, -\Delta + 2\sqrt{U \Delta} )$ interval, which occurs with a  probability $\sqrt{U/\Delta}$.
Superfluidity requires at least that all occupied sites form a percolating cluster, hence we can put the transition line at
\begin{equation}
\frac{U}{\Delta} \gtrsim \frac{1}{p_c^2},
\end{equation}
with $p_c$ the percolation threshold which is $p_c \approx 0.31$ \cite{Isichenko1992} for a simple cubic 3d lattice.
This estimate is in good quantitative agreement with the Monte Carlo results shown in Fig.~\ref{fig:phasediagram_disorder} for intermediate interaction strengths $U/t \le U_c/t = 29.34(2)$.
We shall see below that similar considerations also hold in lower dimensions.

\subsection{Disordered Bose-Hubbard model in 2d}
\label{sec:2d}

\begin{figure}
\begin{center}
\includegraphics[width=0.8\columnwidth]{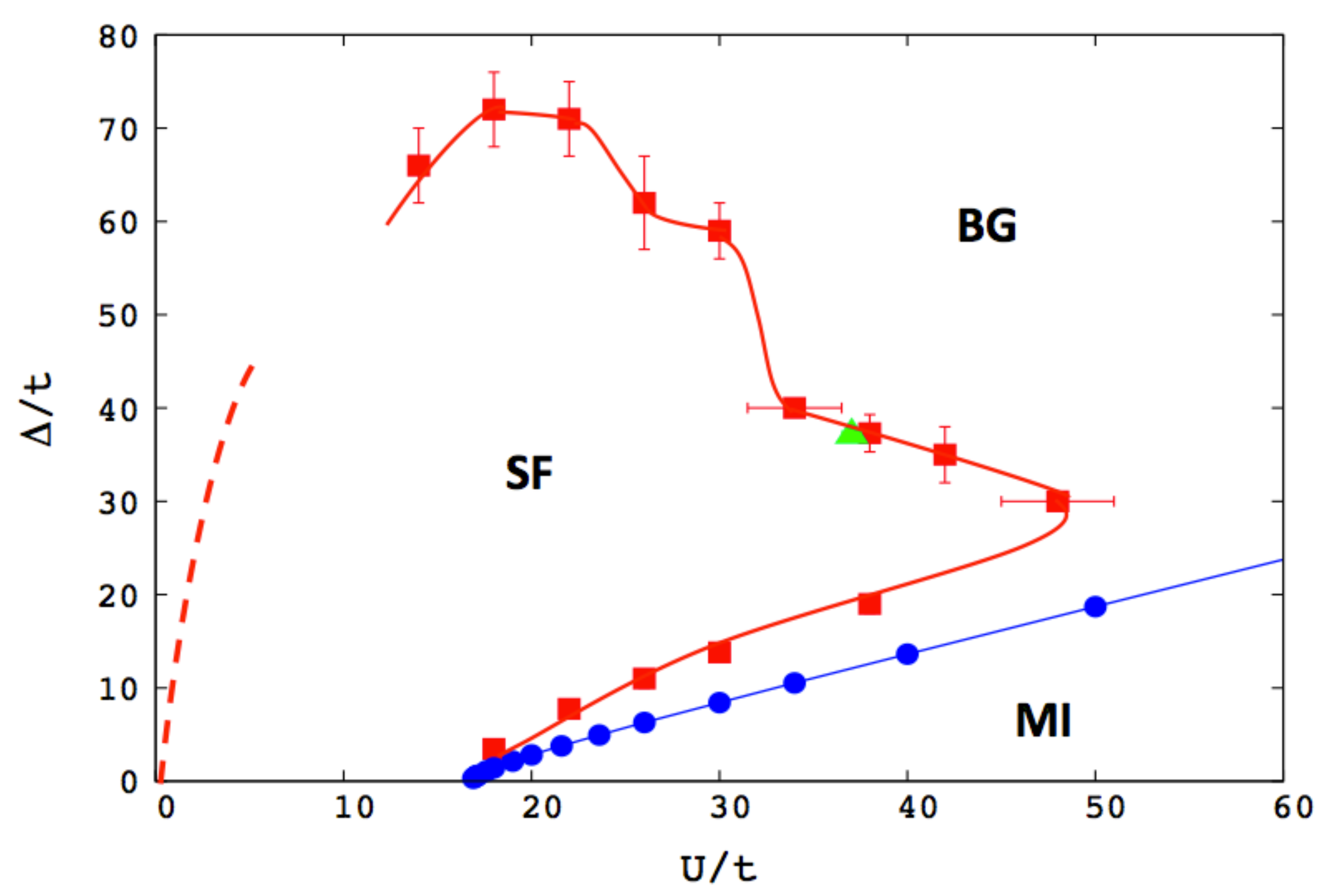}
\caption{ (Color online). Zero temperature phase diagram of the disordered 2d Bose-Hubbard model at unit filling. The MI-BG transition at $\Delta = E_{g/2}$ is obtained using the gap data from Ref.~\cite{CapogrossoSansone_3d}. The green triangle is the point on the SF-BG boundary obtained in Ref.~\cite{Lin2011}.
Reprinted figure with permission from Ref.~\cite{Soyler2011}. Copyright (2011) by the American Physical Society.
}
\label{fig:phasediagram_disorder_2d}
\end{center}
\end{figure}

\begin{figure}
\begin{center}
\includegraphics[width=0.8\columnwidth]{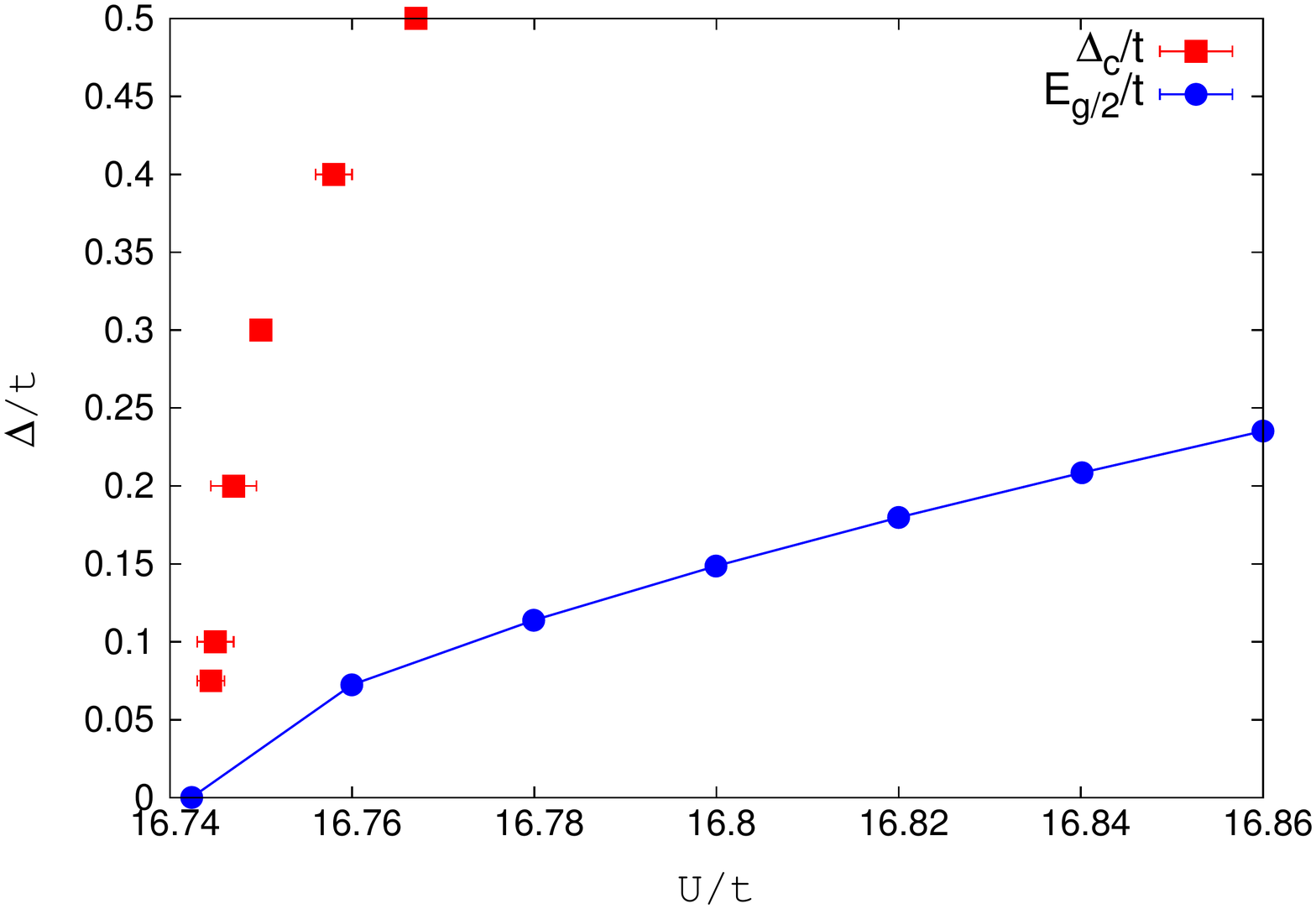}
\caption{ (Color online). Zero temperature phase diagram of the disordered 2d Bose-Hubbard model at unit filling in the vicinity of the tip of the Mott lobe. The MI-BG transition at $\Delta = E_{g/2}$ is obtained using the gap data from Ref.~\cite{CapogrossoSansone_3d}. Note the extremely weak dependence of the SF-BG critical line on disorder for small $\Delta$.
Reprinted figure with permission from Ref.~\cite{Soyler2011}. Copyright (2011) by the American Physical Society.
}
\label{fig:phasediagram_disorder_2d_critical}
\end{center}
\end{figure}

In two dimensions, the topology of the phase diagram is the same as in 3d (see Fig.~\ref{fig:phasediagram_disorder_2d})~\cite{Soyler2011}. Reentrant behavior is seen, there is also a superfluid "finger", and for intermediate interactions the superfluid survives up to disorder strengths rougly 10 times the bandwidth, $\Delta/t = 72$. The finite temperature phase transition between the superfluid and the normal phase is of Kosterlitz-Thouless type with a very strong system size dependence. The superfluid is hence very fragile against temperature fluctuations but robust against strong interactions and disorder. Interestingly, the authors of Ref.~\cite{Soyler2011} report that they were unable to make an unambiguous case for the form of the transition line $U_c(\Delta)$ in the vicinity of the tip of the Mott lobe (see Fig.~\ref{fig:phasediagram_disorder_2d_critical}). For very small $\Delta$, the first critical points could not be distinguished from the critical value in the clean system at $U_c(\Delta = 0) = 16.7424(5)$ despite the very low error bars. The authors argue that disorder becomes relevant on length scales $L > (\kappa \Delta)^{-2/d}$, but this implies an exponentially large system size in 2d, $L > \xi > \exp(U / \Delta)$ when the phase of a vortex loop using Popov's hydrodynamic action, is estimated.
The SF-BG transition line is hence determined by the non-universal microscopic physics. Only in 1d does disorder determine the shape of the transition line~\cite{Svistunov1996}.

\begin{figure}
\begin{center}
\includegraphics[width=0.8\columnwidth]{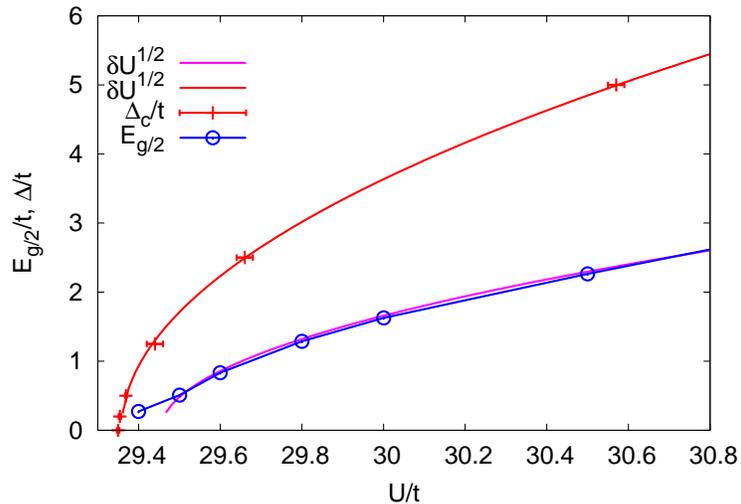}
\caption{ (Color online). Critical disorder bound $\Delta_c /t$ near the tip to the Mott insulator lobe for the 3d case. Also shown are the half gaps $E_{g/2}$ for the disorder free system taken from Ref.~\cite{CapogrossoSansone_3d}. The fits through the data are square roots. For $\Delta_c$ the fit extends at least till $\Delta = 10$. For the disorder free system, the mean-field exponent $\nu = 1/2$ fits through the data only for relatively small gaps, $E_{g/2} \le 2t$. Close to the tip of the lobe, the gaps deviate from the fit, possibly due to the logarithmic corrections to scaling (the quantum critical point is exactly at the uppper critical dimension). The fits for both quantities extrapolated to $\Delta = E_{g/2}=0$ yield a small difference in the location of the critical point.
}
\label{fig:neartip_3d}
\end{center}
\end{figure}

These arguments imply that also in 3d we expect a vertical slope at $U_c$.  It is seen in Fig.~\ref{fig:neartip_3d} that $\Delta_c$ is indeed insensitive to small changes in $U- U_c(\Delta=0) = \delta U$, and the fit suggests $\Delta_c \propto (\delta U)^{1/2}$ (which is the most natural situation) and hence a vertical slope at $U_c$. The critical line $\Delta_c(U)$ has been determined on the basis of winding number fluctuations. To this end, we needed a more accurate determination of the tip of the lobe in the disorder free case, namely $U_c = 29.350(5)$  (determined on the basis of winding number fluctuations for system sizes up to $L=48$), in agreement with the previously published result $U_c = 29.34(2)$~\cite{CapogrossoSansone_3d} (determined on the basis of the gaps in the MI for system sizes up to $L=20$). It is interesting to note in Fig.~\ref{fig:neartip_3d} that the extrapolation of the gaps in the pure MI and the extrapolation of the SF-BG transition points in case of disorder, lead to a small difference in the location of the critical point. This may be related to logarithmic corrections to scaling in the disorder free case, where we are exactly at the upper critical dimension. 

In Ref.~\cite{Lin2011} the dynamic conductivity as a function of Matsubara frequency for a specific point on the SF-BG transition line in 2d was computed (see the point on the line $\Delta = U$ indicated by a triangle in Fig.~\ref{fig:phasediagram_disorder_2d}). The universal conductivity in the high-frequency limit~\cite{Damle97} was estimated to be around $0.17 \sigma_Q$ with $\sigma_Q = e^{*2}/ \hbar^2$ the quantum conductivity unit expressed in terms of the effective charge $e^*$ of the bosons. This value was in agreement with previous calculations on quantum rotor models~\cite{Sorensen92}. The dynamic conductivity shows however clear deviations from scaling with $\omega$, consistent instead with the expected scaling with $\omega /T$.

\subsection{Phase diagram of the disordered Bose-Hubbard model in 1d}
\label{sec:1d}

\begin{figure}
\begin{center}
\includegraphics[width=0.8\columnwidth]{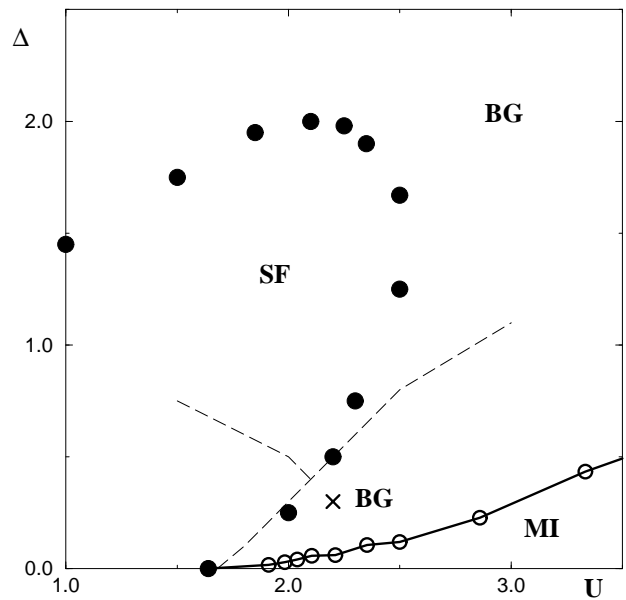}
\caption{ (Color online). Phase diagram of the disordered 1d Bose-Hubbard model at unit filling. For comparison with the units used in this paper, the values of $U$ and $\Delta$ need to be multiplied by 2 in the figure. Error bars are smaller than the pointsize. The BG phase was verified for a system of size $L=1000$ at the point marked by a cross. The dashed line indicates results obtained in Ref.~\cite{Pai1996} in violation of Sec.~\ref{sec:theorem_Fisher}.
Reprinted figure with permission from Ref.~\cite{Prokofev1998_comment}. Copyright (1998) by the American Physical Society.
}
\label{fig:phasediagram_disorder_1d}
\end{center}
\end{figure}

In one dimension, Svistunov could show 10 years before the theorem of inclusions was known that the BG phase must intervene between the MI and SF~\cite{Svistunov1996}. The phase diagram was computed by Monte Carlo simulations~\cite{Prokofev1998_comment} (shown in Fig.~\ref{fig:phasediagram_disorder_1d}) and by the density matrix renormalization group~\cite{Rapsch1999} (the DMRG data in Fig.~\ref{fig:phasediagram_disorder_1d} are not the ones of Ref.~\cite{Rapsch1999}, whose data are not in violation of Sec.~\ref{sec:theorem_Fisher} and are closer to the Monte Carlo data, but the agreement is rather  qualitative).

Interestingly, controversy remains about the nature of the SF-BG transition in one dimension. Giamarchi and Schulz performed a lowest order RG scaling and found that the 
transition is of Kosterlitz-Thouless type with the Luttinger parameter taking an universal value $K = 3/2$  (instead of $K=2$ for the SF-MI transition in the clean system)~\cite{Giamarchi1988}.  In Refs.~\cite{Kashurnikov1995, Svistunov1996} this was shown to hold for any finite disorder by invoking a picture of the unbinding of instanton--anti-instanton pairs. A recent calculation to two loops in the RG showed that the Luttinger parameter remains unchanged at the transition and also correlation functions keep universal exponents as long as bosonization applies~\cite{Ristivojevic13}. However, speculation remained about a possible strong-disorder scenario that would be disorder-driven instead of interaction-driven. In the scenarios put forward in Refs.~\cite{Altman2004, Altman2008, Altman2010} the transition for $\Delta/U \gg 1$ is believed to be still of Kosterlitz-Thouless type, but with strong power law distributions of the Luttinger parameter, whose average is then non-universal on the transition line. This RG procedure is also explained in the paper by Refael and Altman published in the same dossier of this journal~\cite{Gil2013}.

The phase diagram shown in Fig.~\ref{fig:phasediagram_disorder_1d} was determined by locating where the Luttinger parameter takes a value $3/2$ on a system size of $L=100$. Close to the MI this is certainly fine and a full Kostertlitz-Thouless scaling analysis will not shift the data points outside error bars. One notices that the transition line has not been determined for $U/t \le 2$ (in the units used in this paper, not in the figure). Indeed, simulations show that the distribution of the Luttinger parameter becomes very wide, making simulations very cumbersome. Very recently new insights revealed that a strong disorder fixed point leads to a prolonged flow of a renormalized classical field where the fugacity of the instanton--anti-instanton pairs is so low that they are ineffective on mesoscopic length scales~\cite{Pollet2013}. It was also shown that self-averaging of the inverse stiffness, formulated in terms of a relative characteristic width determined by the percentiles of the distribution, always takes place in the superfluid phase and on the critical line (even if the variance would diverge), implying that the quantum transition in the thermodynamic limit always occurs for $K=3/2$ but it may be masked by a logarithmically slow classical flow on all numerically accessible system sizes. This theory was illustrated for a J-current model~\cite{Wallin1994} in (1+1)d in the vicinity of a first-order transition point in the disorder-free case (which speeds up the calculations substantially compared to the quantum models), but all the arguments equally apply to the 1d Bose-Hubbard model. Such a calculation has not been completed yet, but an analysis along the lines of Ref.~\cite{Pollet2013} can now be done at moderate computational cost. The implications are that $K=3/2$ holds everywhere on the SF-BG line in the thermodynamic limit (thus contrasting Ref.~\cite{Gil2013}) but that the mesoscopic flow may be of greater physical relevance, and that the law Eq.~\ref{eq:lowU} also applies for weak $U$.

\section{Experimental systems}
\label{sec:experiment}

Bosons with disorder can be realized in a number of different ways. 
They include Cooper pairs in thin superconducting films ~\cite{Goldman1995}, Josephson junction arrays~\cite{Zant1996}, and $^4$He in porous media~\cite{Crowell97} and on substrates~\cite{Csathy2003}. The detection of the Bose glass phase in these systems has remained ambiguous however. The experiments on $^4$He in porous media for example were more convincingly explained by a model which has a constant density of states for low and for high energies, with a gap in between~\cite{Crowell97}.
Recently, also cold atom systems came at our disposal, with their unique properties of tunability and control.
 In the experiments of Ref.~\cite{White09, Pasienski09} optical speckles were used to generate the disorder. It was not possible to distinguish between a Mott insulator and a Bose glass phase, only insulating phases could be distinguished from superfluid phases on the basis of time-of-flight interference images and transport measurements. They found insulating phases for disorder strengths several hundred times the tunneling amplitude, in agreement with the quantum Monte Carlo simulations for $U < U_c(\Delta=0)$ in Fig.~\ref{fig:phasediagram_disorder}. However, they saw no signs of disorder induced superfluidity (missing the 'finger' in Fig.~\ref{fig:phasediagram_disorder}). Although the disorder distributions are different in experiment and in simulations (see Ref.~\cite{Zhou2010} for more realistic system parameters), the topology of the phase diagram should be the same. The discrepancy is attributed to the low transition temperature (or equivalently, the low superfluid density at zero temperature in the finger~\cite{Gurarie2009}) while the temperature in experiment is estimated to be well above it. Finite system sizes (they are currently not larger than what can be done numerically) also play a role, certainly when only a single disorder realization is used (as is currently the case). However, localization due to disorder but in the absence of interactions was successful~\cite{Billy2008} as well as in the context of the Aubry-Andre model~\cite{Roati2008}.

Finally, we wish to mention that there have been a number of recent experiments on disordered quantum antiferromagnets~\cite{Yamada09, Hong09, Yu11}, supported by simulations in Ref.~\cite{Roscilde2007}. These systems are, however, not without criticism either~\cite{Zheludev09, Wulf11} and we refer to the paper by Zheludev and Roscilde that is published in the same dossier of this journal for a detailed discussion~\cite{Zheludev13}.

\section{Significant others}
\label{sec:other}

There have been many interesting studies on disordered bosonic systems that are omitted here for reasons of space. We mention just a few examples.

Disordered bosonic systems in continuous space were studied in Refs.~\cite{Pilati_disorder1, Pilati_disorder2}, where localization effects due to interaction (MI) are absent. The authors looked at the suppression of $T_c$ of the SF caused by disorder, modeled by an isotropic 3d speckle potential. Unlike the disorder-free system $T_c$ changed considerably between $na^3 = 10^{-4}$ and $na^3 = 10^{-6}$ (An older study found no substantial drop in $T_c$~\cite{Gordillo00}). Agreement with a  perturbative approach for $\delta$-correlated disorder could not unambiguously be established since the precision in the weak disorder limit was not high enough.
For strong interactions and low enough temperature, a normal phase with energy scaling as $\sim T^2$ was found, compatible with a BG at $T=0$.

Instead of a microscopic quantum system, there have been simulations on classical J-current systems ~\cite{Wallin1994} in a higher dimension which have the same universal physics near the transitions but different microscopics unrelated to the Bose-Hubbard model. An older study ~\cite{Balabanyan2005} in 1d found a Kosterlitz-Thouless transition compatible with a universal Luttinger value (cf.~\cite{Giamarchi1987, Giamarchi1988}), but the disorder was too weak to see the physics mentioned in Ref.~\cite{Pollet2013}. In 2d, the $z=d=2$ universality class on the SF-BG transition line could be established for diagonal disorder~\cite{Prokofev2004} with a finite and non-singular compressibility, but  the universal behavior sets in only at very large space-time distances. The authors looked also at off-diagonal disorder, where the symmetry is different, resulting in the vanishing of the compressibility of the SF-BG transition line (the BG phase remains gapless) and a dynamical exponent numerically determined as $z=1.5(2)$.

\section{Conclusion}
\label{sec:conclusion}
Following up on studies of Helium-4 in porous media in the 80s, Giamarchi and Schulz introduced the Bose-Glass phase in a disordered Bose-Hubbard Hamiltonian in one dimension~\cite{Giamarchi1987, Giamarchi1988}. Fisher {\it et al.} argued the existence of this phase in any dimension and provided an in-depth analysis~\cite{Fisher1989}. The model has remained subject of debate ever since, even at the qualitative level. Especially the relevance of weak disorder at the SF-MI transition line remained controversial. Over the last 5 years however, a number of analytical arguments and numerical simulations have been presented which result in a fairly complete phase diagrams of the disordered Bose-Hubbard model in three~\cite{Gurarie2009} and two dimensions~\cite{Soyler2011}, with a final verdict on the issue whether a direct transition between the SF and the MI is possible in the presence of disorder (it is not). An analytical understanding of the shape of the superfluid 'finger' is still lacking, and also the shape of the SF-BG transition line near the tip of the Mott lobe requires further understanding. Even in one dimension, the completion of the phase diagram along the lines of Ref.~\cite{Pollet2013} now seems possible. With the help of Monte Carlo simulations, we have thus arrived at a comprehensive understanding of the thermodynamic properties of the disordered Bose-Hubbard system, both on mesoscopic scales as well as in the thermodynamic limit.

I am grateful to my disorder collaborators V. Gurarie, N. Prokof'ev, B. Svistunov and M. Troyer for countless valuable discussions and insights. 
This work is supported by the Excellence Cluster NIM, FP7/Marie-Curie Grant No. 321918 ("FDIAGMC") and FP7/ERC Starting Grant No. 306897 ("QUSIMGAS").








\end{document}